# Defining Data Science

*Beyond the study of the rules of the natural world as reflected by data*

Yangyong Zhu and Yun Xiong

School of Computer Science, Fudan University, Shanghai, China

Shanghai Key Laboratory of Data Science, Fudan University, China.

{yyzhu, yunx}@fudan.edu.cn

Data science has received widespread attention in academic and industrial circles. New data science research institutes and organizations have continued to emerge on the scene, such as the Columbia University Institute for Data Sciences and Engineering and New York University Center for Data Science. The University of California at Berkeley, Columbia University, Fudan University, and other universities have launched data science courses and degree programs. Cleveland and Smith proposed that data science should be considered an independent discipline[2, 8]. Facebook, Google, EMC, IBM, and other companies have established employment positions for data scientists. According to *Harvard Business Review*, the data scientist is "the sexiest job of the 21st century." Currently, there are several viewpoints regarding the definition of data science (see page 2). However, there is no consensus definition. We believe that, as a new science, the research objectives of data science are different from those of other, more established branches of science. In addition, the scientific issues that data science addresses are not studied by natural or social sciences.

Our team has worked on data technology and research projects funded by Chinese government since 1998, and we have applied our work to life science, healthcare, finance, transportation, and other fields (Table 1). Over the years, we have noticed a number of common issues related to data in scientific research and industrial applications, most notably the similarity of data objects. We have come to realize that there is a considerable need to conduct research specifically on the data itself, and we started to explore concept of data science in 2009[9]. Since 2010, we have hosted the annual International Symposium on Data Science and Dataology (iwdds.fudan.edu.cn). The symposium provides us with forum for the discussion of data science issues with scientists involved in computer science, life sciences, astronomy, and other fields. Over the past 16 years, our understanding of data science has taken more solid shape. We believe that data in cyberspace have formed what we call datanature[9, 10]. Data science is the scientific research of datanature.

There are several current viewpoints on data science.

**VP1: Data science is the science of studying scientific data.**

The Committee on Data for Science and Technology (CODATA) launched the *Data Science Journal* (codata.org/dsj/) in 2002. CODATA regards data science as the methods and technologies used to conduct scientific research through management and utilization of scientific data. As scientific data have become more accessible, data science has been used to better characterize the data-intensive nature of today's science and engineering. Many disciplines use data technology to deal with scientific data from their respective areas. From this, X-informatics emerged, including bioinformatics, neuroinformatics, and social informatics.

For example, researchers in NuMedii, Inc., a big-data company in Silicon Valley, predicted whether existing drugs could be used to treat ovarian cancer by examining gene expression data from over 2,500 ovarian tumor samples[6].

As another example, mathematicians from Harvard University Aiden and Michel studied American history using Ngrams on Google[1]. They used Ngrams to search for the usage frequencies of two phrases: "United States are" and "United States is." The search results showed that before the American Civil War, the two phrases were used at roughly equal frequency, but after the Civil War, the latter became far more common than the former. This is seen as indicative of the levels of acceptance by the public of the United States as a unified nation before and after the Civil War.

From this point of view, data mainly refer to data generated and used in scientific studies. This emphasizes that data science is the management, processing, and use of scientific data to support scientific research, i.e., the currently commonly known data-intensive scientific research or fourth paradigm of scientific research[4].

**VP2: Data science is the science of studying business data.**

In 2010, Loukides discussed what data science is, arguing that data science should enable the creation of data products rather than working as a simple application with data[5]. In 2013, Provost et al. pointed out, "extracting knowledge from data to solve business problems" is one of the fundamental concepts of data science[7].

Providing support for BI methodology research makes up a significant portion of the work performed by many data scientists. To effect this, a large proportion of BI practitioners were transitioned into data scientists. Amazon, Google, LinkedIn, Facebook, and other internet companies opened job positions for data scientists and established data science teams. These data scientists study and analyze business data to provide services for management decision making. For example, Amazon uses collaborative filtering to generate high-quality product recommendations, and Facebook uses a "People you may know" feature to recommend friend connections.

From this point of view, the acquisition of knowledge from business data in order to make decisions is one aspect of data science. This is similar to what BI scientists work on. For this reason, many BI scientists are also called data scientists. However, compared to BI issues, data science focuses more on common issues in the analysis of various business data, i.e., the issues on BI methodology.

**VP3: Data science is an integration of statistics, computing technology, and artificial intelligence (AI).**

This viewpoint often comes up in discussions on what data scientists are. It is generally believed that data scientists should have skills in statistics, computing technology, AI, and related fields and that data scientists are not individual people specializing in one field so much as teams consisting of statisticians, computer scientists, AI experts, and domain experts.

For example, the data scientist teams at Google and Facebook are composed of statisticians, computer scientists, AI scientists, and experts in other relevant fields.

This viewpoint is simple: Because statistics, computing technology, and AI are all used to process and analyze data, they are all a natural part of data science.

**VP4: The purpose of data science is to solve scientific and business problems by extracting knowledge from data.**

In 2013, Dhar discussed the implications of data science from a business and research standpoint[3]. He defined data science as "the study of the generalizable extraction of knowledge from data"[3]. He also pointed out that a data scientist needs to have comprehensive skills covering statistics, machine learning, AI, and database management and have a deep understanding of problem design. This viewpoint can be seen as an integration of the first three viewpoints.

**What is data science?**

The basic ideas underlying the definitions given above are that data science is used to acquire knowledge from data in some relevant fields and to provide support for existing scientific research and management decision-making schema. However, all work described above is still not enough to establish data science as a new, unique branch of science. This is because the objects of their study are things in the natural world, and their research issues are also addressed in existing scientific fields.

With the development of digital equipment, things in the natural world are increasingly being stored in cyberspace in the form of data. Data are entered, generated, and created in cyberspace in a variety of ways and have become more and more diverse, complex, and out of human control. More and more data are unknown to or poorly understood by humans. Data in the cyberspace already show features of an independent world, like the natural world, so all data in cyberspace are here referred to as datanature.

It should be noted that there are two types of data in the cyberspace. The first is the data that represent things in the natural world, here called real data. An example is personal information, which is data representative of personal characteristics. The second is data that do not represent things in the natural world, here called virtual data. Virtual data means that the instances of such data have no references in the natural world. An example is computer viruses, which are neither viruses in the natural world nor data representation of real viruses; instead, they only exist in cyberspace (Figure 1).

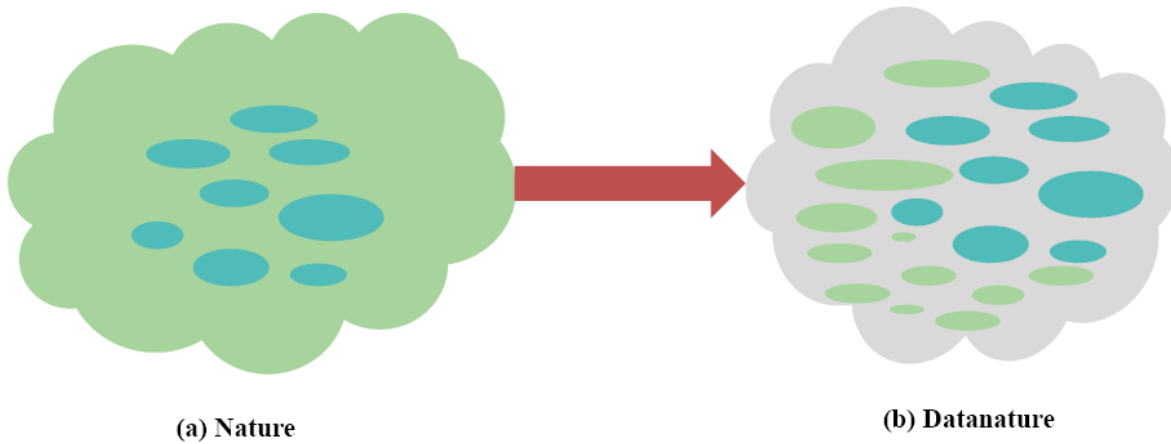

**Figure 1. Real data and virtual data.** The blue ellipses and the green ellipses in Figure 1(b) denote the real data that represent things in the natural word. Among these, the blue ellipses correspond to those things in the natural world which have been stored in cyberspace in the form of data (i.e., the blue ellipses in Figure 1(a)); the green ellipses correspond to those things in the natural world which would be stored in cyberspace gradually (i.e., the green part in Figure 1 (a)). The grey part in Figure 1(b) denotes the virtual data that do not represent things in the natural world, i.e., the instances of such data have no references in the natural world.

The formation of datanature has produced new objects of study and new scientific issues. These new objects of study are not things that exist in the natural world or in human society but rather in datanature, i.e., data. There are new scientific issues about datanature. What size is datanature? What is the growth rate of global data? How do the data flow in cyberspace? How should the authenticity of datanature be determined? None of these issues are addressed by the natural or social sciences. These new scientific issues need to be studied by a new science.

Data science is the science of studying datanature and the science of data itself. On a basic level, it involves extracting knowledge from data. Because some of the data in datanature do represent real things, the knowledge acquired from these data can be used for natural and social science. This type of work is considered as data science according to VP1-VP4. However, it is only one part of data science research.

Data science is here defined as follows:

Data science is the theory, method, and technology of studying datanature. It has two main components.

**The first component is the study of the patterns and rules of data itself. Its goal is to explore datanature and scientific issues related to datanature.** This does not take into account

the meaning of the data in the natural world.

**The second component is the study of the rules of the natural world as reflected by data, i.e., the study the natural world performed through the study of data.** For example, the purpose of performing a study on data representing a person's behavior is to study that person's behavior.

As mentioned above, studies on data have been under way for some time, and data techniques such as data mining have been developed. However, the data science research community needs to establish fundamental theory and basic methods for scientific observation and measurement and to further develop data techniques. Figure 2 shows the main topics of data science.

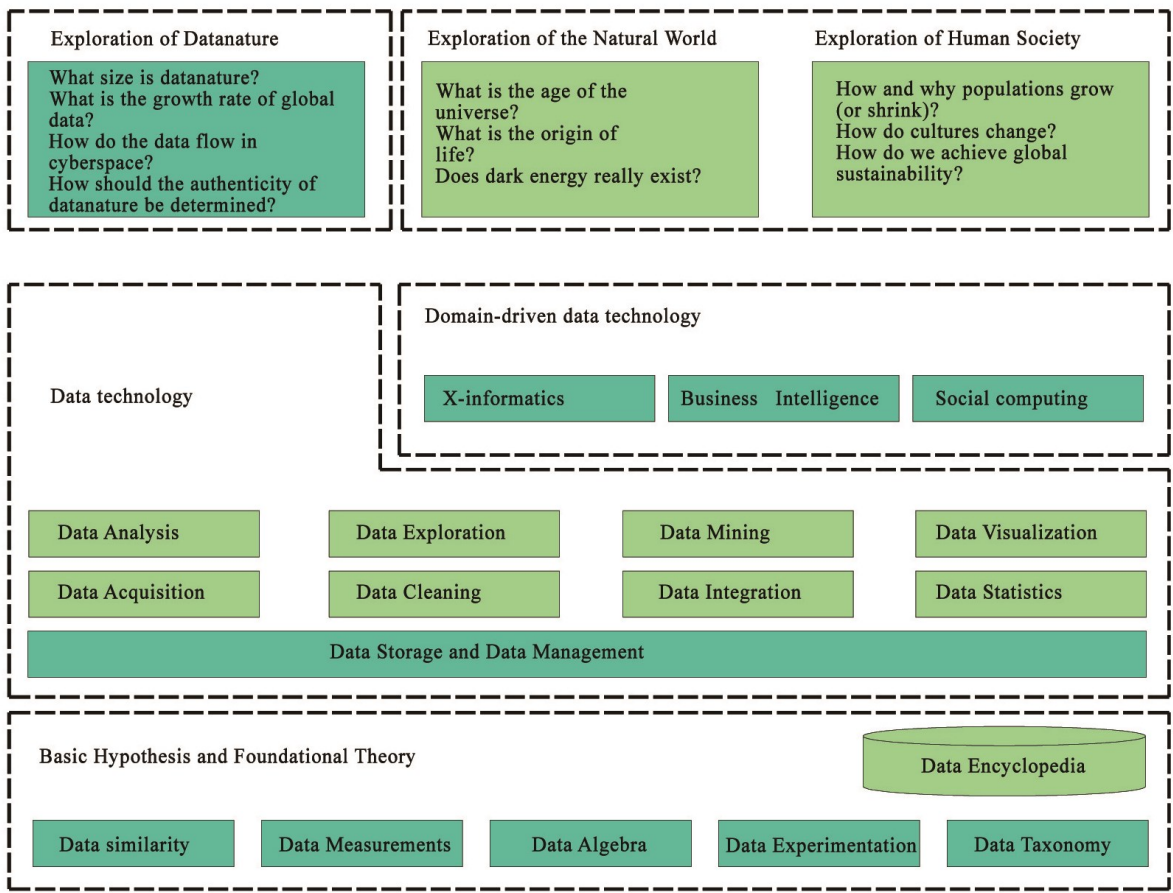

Figure 2. Research Topics of Data Science

## Conclusions

Data science is gaining more and more and widespread attention, but no consensus viewpoint on what data science is has emerged. As a new science, its objects of study and scientific issues should not be covered by established sciences. In the present paper, data science is defined as the science of exploring datanature. We believe this is the most logical and accurate definition of data science, and it includes key parts of definitions VP1-VP4.

**Reference.**

**Yangyong Zhu** (yyzhu@fudan.edu.cn) is the director of the Shanghai Key Laboratory of Data Science, Fudan University, China and a professor at the School of Computer Science, Fudan University, Shanghai, China.

**Yun Xiong** (yunx@fudan.edu.cn) is an associate professor at the School of Computer Science, Fudan University, Shanghai, China and a data scientist at the Shanghai Key Laboratory of Data Science, Fudan University, China.

Table 1. Research projects in data fields in which our team participated.

| Id | Year | Project name | Funding source |
|---|---|---|---|
| 1 | 1998 | Development and application of tools for data mining | National High Technology Research and Development Program of China (863 Program) |
| 2 | 2000 | Development and application of bioinformatics platform for drug target identification | National High Technology Research and Development Program of China (863 Program) |
| 3 | 2002 | Development and application of data mining platform | National High Technology Research and Development Program of China (863 Program) |
| 4 | 2004 | Optimization technology and implementation of Shanghai World Expo Website | Shanghai Municipal Government |
| 5 | 2005 | Research on data model, index and architecture of biological sequence database | National Natural Science Foundation of China |
| 6 | 2006 | Mining of transcription factor binding sites and composite modules | National High Technology Research and Development Program of China (863 Program) |
| 7 | 2008 | Research on key technologies in data analysis for medical-insurance-fund risk prevention | Shanghai Municipal Government |
| 8 | 2009 | A new generation of high-definition images for intelligent traffic monitoring and industrialization of the information service system | Shanghai Municipal Government |
| 9 | 2010 | Research on feature-based mining algorithm to build biological networks | National Natural Science Foundation of China |
| 10 | 2011 | Development and operations of cloud platform for the stock market | National Science and Technology Support Program |
| 11 | 2012 | Research on mining algorithm for peculiar groups | National Natural Science Foundation of China |
| 12 | 2012 | Research on medical big data | Shanghai Municipal Government |
| 13 | 2012 | Theoretical research on urban traffic big data | Shanghai Municipal Government |
| 14 | 2013 | Research on the theories of data science | Shanghai Municipal Government |
| 15 | 2014 | Research on big data technology for Shanghai shipping industry | Shanghai Municipal Government |